\documentclass[11pt]{article}

\usepackage{graphicx,amssymb,amsbsy}
\usepackage{natbib}

\newcommand\Real{\mbox{Re}} 

\newsavebox{\astrutbox}
\sbox{\astrutbox}{\rule[-5pt]{0pt}{20pt}}

\def\pr{{\partial}}

\begin{document}

\centerline{\Large{\bf Instability of a viscous flow between rotating porous}}
\centerline{\Large{\bf  cylinders with radial flow}}

\vskip 5mm
\centerline{\bf Konstantin Ilin\footnote{Department of Mathematics, University of York,
Heslington, York YO10 5DD, UK. Email address for correspondence: konstantin.ilin@york.ac.uk} and
Andrey Morgulis\footnote{Department of Mathematics, Mechanics and Computer Science, The Southern Federal University, Rostov-on-Don, and South Mathematical Institute, Vladikavkaz Center of RAS, Vladikavkaz,
Russian Federation}}

\begin{abstract}
\noindent
The stability of a two-dimensional viscous flow between two rotating porous cylinders is studied.
The basic steady flow is the most general rotationally-invariant solution of the Navier-Stokes
equations in which the velocity has both radial and azimuthal components, and the azimuthal velocity profile
depends on the Reynolds number. It is shown that for a wide range of the parameters of the problem,
the basic flow is unstable to small two-dimensional perturbations. Neutral curves in the space of parameters of the problem are computed.
Calculations show that the stability properties of this flow are determined by the azimuthal velocity at the inner cylinder
when the direction of the radial flow is from the inner cylinder to the outer one and by the azimuthal velocity at the outer cylinder
when the direction of the radial flow is reversed. This work is a continuation of our previous study of an inviscid instability in flows between
rotating porous cylinders (see \cite{IM2013}).
\end{abstract}


\setcounter{section}{0}
\setcounter{subsection}{0}


\setcounter{equation}{0}
\renewcommand{\theequation}{1.\arabic{equation}}


\section{Introduction}

In \cite{IM2013}, we have presented an example of a steady inviscid flow in an annulus between
two rotating porous cylinders which is unstable to small two-dimensional perturbations in the framework of the inviscid theory.
The aim of the present paper is to show that this instability occurs not only for a particular steady flow considered in \cite{IM2013}, but for the whole class of steady rotationally-symmetric viscous flows between
two rotating porous cylinders.

The basic steady flow represents the most general rotationally-invariant solution of the Navier-Stokes
equations. In this flow, the velocity has both radial and azimuthal components, and the azimuthal velocity profile
non-trivially depends on the Reynolds number based on the radial velocity. We formulate a linear stability problem,
solve it numerically and compute neutral curves that
separate regions of stability and instability in the space of parameters of the problem.

One of the most interesting features of the instability is that it occurs not only in the diverging flow
(where the direction of radial flow is from the inner cylinder to
the outer one) but also in the converging flow (where the direction of radial flow is reversed).
This contradicts a view that, while diverging flows may be unstable, converging
flows are always stable, which originates in the studies of the Jeffery-Hamel flow (e.g. \cite{Goldshtik1991,Shtern,Drazin1998}).
Another interesting property of this instability is that it is almost unaffected by what is happening at the flow outlet
(i.e. at the outer cylinder for diverging flows and at the inner cylinder for converging flows): the stability properties of the flow
are determined by the azimuthal velocity at the flow inlet and the effect of inlet conditions remains strong in the whole flow domain.
This is somehow in contrast with viscous flows through channels and pipes where the effect of the inlet conditions weakens with
distance from the inlet due to viscosity.

The stability of viscous flows between permeable rotating cylinders with a radial flow to three-dimensional perturbations
had been studied by many authors
(e.g. \cite{Bahl, Chang, Min, Kolyshkin, Kolesov}).  One of the main aims of these studies was to determine the effect of a radial flow
on the stability
of the circular Couette-Taylor flow, and the general conclusion was that it changes the stability properties of the flow:
both a converging radial flow and a sufficiently strong diverging flow have a stabilizing effect on the Taylor instability,
but when a divergent flow is weak, it has a destabilizing effect \citep{Min, Kolyshkin}. However, it remained unclear whether
a radial flow itself can induce instability for flows which are stable without it. The present paper and results of \cite{IM2013}
give an answer to this question.

The outline of the paper is as follows. In Section 2, we formulate the problem and describe the basic steady flow. Section 3 contains a linear inviscid stability analysis. Conclusions are presented in Section 4.


\setcounter{equation}{0}
\renewcommand{\theequation}{2.\arabic{equation}}


\section{Formulation of the problem and basic steady flow}\label{sec:problem}
Consider two-dimensional viscous incompressible flows in an annulus between two concentric porous cylinders
with radii $r_{1}$ and $r_{2}$ ($r_2 > r_1$). We assume that there is a constant volume flux $2\pi Q$ of
the fluid across the annulus (through the walls of the cylinders). $Q$ is positive if the direction of the fluid flow is
from the inner cylinder to the outer one
and negative if the fluid flows in the opposite direction. Flows with $Q>0$ will be referred to as diverging flows, while
flows with $Q<0$ will be called converging flows.
We take $r_1$ as a length scale, $r^2_{1}/\vert Q\vert$ as a time scale, $\vert Q\vert/r_{1}$ as a scale for the velocity and $\rho Q^2/r_{1}^2$ for the pressure where $\rho$ is the (constant) fluid density. Then the two-dimensional Navier-Stokes equations,
written in non-dimensional variables, have the form
\begin{eqnarray}
&&u_{t}+ u u_{r} + \frac{v}{r}u_{\theta} -\frac{v^2}{r}= -p_{r} +
\frac{1}{R} \left(\nabla^2 u-\frac{u}{r^2}-\frac{2}{r^2}v_{\theta}\right) ,  \label{1} \\
&&v_{t}+ u v_{r} + \frac{v}{r}v_{\theta} +\frac{u v}{r}= -\frac{1}{r} \, p_{\theta} +
\frac{1}{R} \left(\nabla^2 v-\frac{v}{r^2}+\frac{2}{r^2}u_{\theta}\right)  ,  \label{2} \\
&&\frac{1}{r}\left(r u\right)_{r} +\frac{1}{r} \, v_{\theta}=0.  \label{3}
\end{eqnarray}
Here $(r,\theta)$ are the polar coordinates, $u$ and $v$ are the radial and azimuthal components of the velocity, $p$ is the pressure,
$R = \vert Q\vert/\nu$ is the Reynolds number  ($\nu$ is the kinematic viscosity of the fluid)
and $\nabla^2$ is the polar form of the Laplace operator:
\[
\nabla^2=\pr_r^2 + \frac{1}{r}\pr_r + \frac{1}{r^2}\pr_\theta^2 .
\]
Both components of the velocity are prescribed at the boundary
\begin{equation}
u\!\bigm\vert_{r=1}=\beta, \quad u\!\bigm\vert_{r=a}=\frac{\beta}{a},
\quad v\!\bigm\vert_{r=1}=\gamma_1, \quad v\!\bigm\vert_{r=a}=\frac{\gamma_2}{a}.  \label{4}
\end{equation}
Here
\[
a=\frac{r_2}{r_1}, \quad \beta=\frac{Q}{\vert Q\vert}, \quad \gamma_1=\frac{\Omega_1 r_1^2}{\vert Q\vert}, \quad
\gamma_2=\frac{\Omega_2 r_2^2}{\vert Q\vert},
\]
with $\Omega_1$ and $\Omega_2$ being the angular velocities of the inner and outer cylinders respectively;
$\beta=1$ and $\beta=-1$ correspond to the diverging
and converging flows. The boundary conditions (\ref{4}) model
conditions on the interface between a fluid and a porous wall (see \cite{Joseph}).

\begin{figure}
\begin{center}
\includegraphics*[height=8cm]{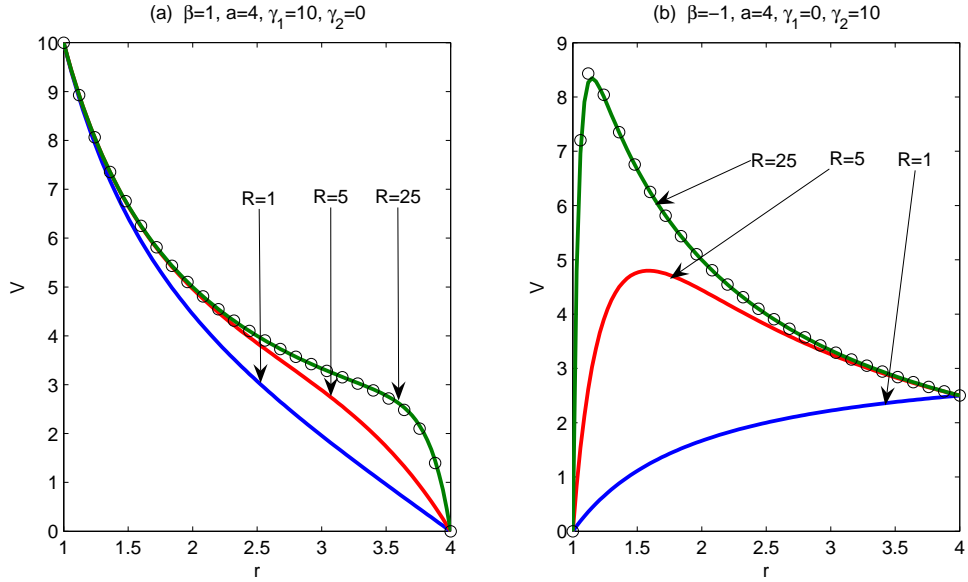}
\end{center}
\caption{Typical velocity profiles for $a=4$ and $R=1,5,25$.
(a) corresponds to the diverging flow ($\beta=1$) with $\gamma_1=10$ and $\gamma_2=0$, (b) corresponds to
the converging flow ($\beta=-1$) with $\gamma_1=0$ and $\gamma_2=10$. Circles show the velocity computed using leading order terms in
the asymptotic formulae (\ref{9}) and (\ref{12}).}
\label{vel_prof}
\end{figure}

Problem (\ref{1})--(\ref{4}) has the following steady rotationally-symmetric solution:
\begin{equation}
u=\frac{\beta}{r}, \quad v= V(r)=A r^{\beta R +1} + \frac{B}{r}  \label{5}
\end{equation}
where
\begin{equation}
A=\frac{\gamma_2 -\gamma_1}{a^{\beta R+ 2}-1}, \quad B=\frac{a^{\beta R + 2}\gamma_1 -\gamma_2}{a^{\beta R + 2}-1}. \label{6}
\end{equation}
The steady solution (\ref{5}) depends on $\gamma_1$, $\gamma_2$ and $\beta R$ and is well defined for all $\beta R\neq -2$.
For $\beta R =-2$, the solution is given by
\begin{equation}
u=-\frac{1}{r}, \quad v= V(r)=\widetilde{A} \, \frac{\ln r}{r} + \frac{\widetilde{B}}{r}  \label{7}
\end{equation}
where
\[
\widetilde{A}=\frac{\gamma_2 -\gamma_1}{\ln a}, \quad \widetilde{B}=\gamma_1.
\]
The dependence of the steady flow (\ref{5}) on $\beta R$ is non-trivial and, for $R \gg 1$, the flow has a boundary layer
either at the outer cylinder (for the diverging flow) or at the inner one (for the converging flow). This is a consequence of the general fact
(which is true for an arbitrary viscous flow in the presence of a non-zero flux of the fluid through a permeable boundary) that
in the limit of high Reynolds number the boundary layer is formed at the outflow part of the boundary
(see, e.g., \cite{Temam,Yudovich2001,Ilin2008}).
Typical velocity profiles $V(r)$
for various $\beta R$ are shown in Fig. \ref{vel_prof}.

Consider the behaviour of the velocity profile when $R\to\infty$ in more detail.
Let $\beta=1$, so that we have the diverging flow.
It follows from Eqs. (\ref{5}) and (\ref{6})  that
\begin{equation}
V(r)=\frac{\gamma_1}{r}+O\left(\left(\frac{r}{a}\right)^{R+2}\right) \quad \textrm{as} \quad R\to\infty. \label{8}
\end{equation}
Therefore, for any $1\leq r\leq\delta< a$, $V(r)$ can be approximated by the inviscid profile $\gamma_1/r$ with exponentially small error.
This approximation fails within the boundary layer of thickness $O(R^{-1})$ near the  outer cylinder. To construct a uniformly valid approximation,
we introduce the boundary layer variable $\eta=R(1-r/a)$ and express the difference $V(r)- \gamma_1/r$ in terms of $\eta$. Then we pass to the limit
as $R\to\infty$ and use the fact that $(1+\alpha/x)^{x}\to e^{\alpha}$ as $x\to\infty$. As a result, we get the following asymptotic formula
\begin{equation}
V(r)=\frac{\gamma_1}{r}+\frac{\gamma_2 - \gamma_1}{a}e^{-\eta} \, \left[1 -\left(\eta+\frac{\eta^2}{2}\right) \, R^{-1}
+O\left(R^{-2}\right)\right]. \label{9}
\end{equation}
The leading term of this formula for $R=25$, $\gamma_1=10$ and $\gamma_2=0$ is shown in Fig. \ref{vel_prof}a. Evidently,
it approximates the exact solution very well.

Note also that everywhere except the boundary layer the flow is irrotational. The vorticity, $\Omega=V'(r)+V(r)/r$, is concentrated within the boundary layer, and the corresponding asymptotic expression is
\begin{equation}
\Omega(r)=  R \, \frac{\gamma_2-\gamma_1}{a^2} \, e^{-\eta} \left[1+O(R^{-1})\right]. \label{10}
\end{equation}

For $\beta=-1$, i.e. for the converging flow, Eqs. (\ref{5}) and (\ref{6})  yield
\begin{equation}
V(r)=\frac{\gamma_2}{r}+O\left(\left(\frac{1}{r}\right)^{R-1}\right) \quad \textrm{as} \quad R\to\infty, \label{11}
\end{equation}
so that, for any $1+\delta\leq r \leq   a$, $V(r)$ can be approximated by the inviscid profile $\gamma_2/r$ with exponentially small error.
Again, this approximation fails within the boundary layer of thickness $O(R^{-1})$ near the inner cylinder, and the uniformly valid approximation
has the form
\begin{equation}
V(r)=\frac{\gamma_2}{r}+(\gamma_1 - \gamma_2)e^{-\xi} \, \left[1 +\left(\xi+\frac{\xi^2}{2}\right) \, R^{-1}
+O\left(R^{-2}\right)\right]. \label{12}
\end{equation}
where $\xi=R(r-1)$. The leading term of (\ref{12}) for $R=25$, $\gamma_1=0$ and $\gamma_2=10$ is shown in Fig. \ref{vel_prof}b and, clearly,
it gives a very good approximation for the exact solution.

The asymptotic formula for the vorticity is
\begin{equation}
\Omega(r)=R \, (\gamma_1-\gamma_2) \, e^{-\xi} \left[1+O(R^{-1})\right]. \label{13}
\end{equation}

In the next section we investigate the stability of the steady flow  (\ref{5}) to small two-dimensional perturbations.


\setcounter{equation}{0}
\renewcommand{\theequation}{3.\arabic{equation}}


\section{Linear stability analysis}\label{sec:stability}

We consider a small perturbation
$(\tilde{u}, \tilde{v}, \tilde{p})$ in the form of the normal mode
\begin{equation}
\{\tilde{u}, \tilde{v}, \tilde{p}\} = Re\left[\{\hat{u}(r), \hat{v}(r), \hat{p}(r)\} e^{\sigma t + in\theta}\right]  \label{3.1}
\end{equation}
where $n\in\mathbb{Z}$. This leads to the eigenvalue problem:
\begin{eqnarray}
&&\left(\sigma +  \frac{in V}{r} + \frac{\beta}{r} \, D \right) \hat{u}
-\frac{\beta}{r^2} \, \hat{u} +\frac{2V}{r} \, \hat{v} = - D \, \hat{p}  +
\frac{1}{R} \left(L \hat{u}-\frac{\hat{u}}{r^2}-\frac{2in}{r^2} \, \hat{v}\right) ,  \label{3.2} \\
&&\left(\sigma +  \frac{in V\gamma}{r} + \frac{\beta}{r} \, D \right) \hat{v}
+\frac{\beta}{r^2} \, \hat{v}  +\Omega(r) u = -\frac{in}{r} \, \hat{p}  +
\frac{1}{R} \left(L \hat{v} -\frac{\hat{v}}{r^2}+\frac{2in}{r^2} \, \hat{u}\right),  \label{3.3} \\
&&D\left(r \hat{u}\right) +in \, \hat{v}=0, \label{3.4} \\
&&\hat{u}(1)=0, \quad \hat{u}(a)=0, \quad \hat{v}(1)=0, \quad \hat{v}(a)=0.  \label{3.5}
\end{eqnarray}
In Eqs. (\ref{3.2}) and (\ref{3.3}),
\[
D=\frac{d}{dr}, \quad L  = \frac{d^2}{dr^2} + \frac{1}{r} \, \frac{d}{dr} - \frac{n^2}{r^2}, \quad \Omega(r)=V'(r)+\frac{V}{r} .
\]
It can be shown that the mode with $n=0$ cannot be unstable (see Appendix A). So, in what follows we consider only the modes with $n\neq 0$.

It is
convenient to introduce the stream function $\hat{\psi}(r)$ such that
\[
\hat{u}(r)=\frac{in}{r} \, \hat{\psi}(r), \quad \hat{v}=-D\hat{\psi}(r).
\]
In terms of $\hat{\psi}(r)$, the eigenvalue problem takes the form
\begin{eqnarray}
&&\left(\sigma +  \frac{in V}{r} + \frac{\beta}{r} \, D \right) L \hat{\psi}-\frac{in}{r} \, D\Omega \, \hat{\psi} = R^{-1}L^2 \hat{\psi} ,  \label{3.6} \\
&&\hat{\psi}(1) =0, \quad \psi'(1)=0, \quad \hat{\psi}(a)=0, \quad \psi'(a)=0.  \label{3.7}
\end{eqnarray}
A couple of important observations can be made just by looking at problem (\ref{3.6}), (\ref{3.7}).
First, it follows from (\ref{3.6}), (\ref{3.7}) and (\ref{5}) that for a given $\beta$, an eigenvalue is a function of five parameters:
$\sigma=\sigma(a, n, \gamma_1, \gamma_2, R)$.  Second, if $\sigma(a, n, \gamma_1, \gamma_2, R)$ is an eigenvalue, then so are
$\bar{\sigma}(a, -n, \gamma_1, \gamma_2, R)$ and $\sigma(a, -n, -\gamma_1, -\gamma_2, R)$. (Here $\bar{\sigma}$ is the complex conjugate of $\sigma$.)
These properties imply that
it suffices to consider
only positive $n$ and $\gamma_1$ (or $\gamma_2$).

The eigenvalue problem (\ref{3.6}), (\ref{3.7}) is solved numerically using pseudospectral methods described, e.g., in \cite{Trefethen}.
Before presenting the results of numerical calculations, we will show that in the limit $R\to\infty$ the eigenvalue problem (\ref{3.6}), (\ref{3.7})
reduces to the inviscid problem studied in \cite{IM2013}.


\subsection{Asymptotic behaviour of eigenvalues for $R\gg 1$}

\subsubsection{Diverging flow ($\beta=1$)}

We assume that the asymptotic expansion has the form
\begin{eqnarray}
&& \sigma=\sigma_{0}+R^{-1}\sigma_{1}+R^{-2}\sigma_{1}+\dots ,  \label{3.8} \\
&& \hat{\psi}=\hat{\psi}_{0}(r)+R^{-1} [\hat{\psi}_{1}(r)+\phi_{0}(\eta)]+R^{-2} [\hat{\psi}_{2}(r)+\phi_{1}(\eta)]+\dots  \label{3.9}
\end{eqnarray}
Here $\eta=R(1-r/a)$ is the boundary layer variable introduced in section 2. Functions $\hat{\psi}_{k}(r)$ ($k=0,1,\dots$) represent
the regular part of the expansion, and $\phi_{k}(\eta)$  ($k=0,1,\dots$) give us boundary layer corrections
to the regular part. We assume that the boundary layer part rapidly decays outside thin boundary layer near
$r=a$:
\begin{equation}
\phi_{k}(\eta) =o(\eta^{-s}) \quad {\rm as} \quad \eta\to\infty ,    \label{3.10}
\end{equation}
for every $s>0$ and for each $k=0,1,\dots$

Substituting asymptotic formulae (\ref{9}) and (\ref{10}) into Eq. (\ref{3.6}), we obtain
\begin{equation}
\left(\sigma +  \frac{in \gamma_1}{r^2} + \frac{in}{r} \,\frac{f(\eta)}{a}
+\frac{1}{r} \, D \right) L \hat{\psi}
-  R^2 \, \frac{in}{r} \, \frac{f(\eta)}{a^3} \, \hat{\psi} = R^{-1}L^2 \hat{\psi}    \label{3.11}
\end{equation}
where
\[
f(\eta)=\left(\gamma_2 - \gamma_1\right) \, e^{-\eta} \left[1+O\left(R^{-1}\right)\right].
\]
Two terms on the left side of (\ref{3.11}) which contain $f(\eta)$ are exponentially small everywhere except the boundary layer
of thickness $O(R^{-1})$ near the outer cylinder. So, the regular part of the expansion is obtained by ignoring these terms,
substituting (\ref{3.8}) and the formula
\begin{equation}
\hat{\psi}=\hat{\psi}_{0}(r)+R^{-1} \hat{\psi}_{1}(r)+R^{-2} \hat{\psi}_{2}(r)+\dots  \label{3.12}
\end{equation}
into Eq. (\ref{3.11}) and then collecting terms of equal powers of $R^{-1}$.
This leads to the sequence of equations
\begin{eqnarray}
&&\left(\sigma_{0} +  \frac{in\gamma_1}{r^2} + \frac{1}{r} \, \pr_{r} \right)L \hat{\psi}_{0}= 0, \label{3.13} \\
&&\left(\sigma_{0} +  \frac{in\gamma_1}{r^2} + \frac{1}{r} \, \pr_{r} \right)L \hat{\psi}_{1}= -\sigma_{1}L \hat{\psi}_{0} +
 L^2 \hat{\psi}_{0}, \quad {\rm etc.} \label{3.14}
\end{eqnarray}
To obtain boundary conditions at $r=1$, we
substitute (\ref{3.9}) into the first two boundary conditions
(\ref{3.7}) and use the fact that functions $\phi_k(\eta)$ at $r=1$ are smaller than any power of $R^{-1}$. This results in
\begin{equation}
\hat{\psi}_{k}(1) = \hat{\psi}'_{k}(1)=0  \label{3.15}
\end{equation}
for each $k=0,1,\dots$ For the boundary conditions at $r=a$, the boundary layer part must be taken into account. So we substitute
(\ref{3.9}) into the last two conditions (\ref{3.7}) and use the relation $\pr_{r}=-R \, a^{-1} \, \pr_{\eta}$. As a result, we get
\begin{eqnarray}
&&\hat{\psi}_{0}(a)=0, \quad \hat{\psi}'_{0}(a)-\phi'_{0}(0)=0, \label{3.16} \\
&&\hat{\psi}_{1}(a)+\phi_{0}(0)=0, \quad \hat{\psi}'_{1}(a)-\phi'_{1}(0)=0, \quad {\rm etc.} \label{3.17}
\end{eqnarray}
Equation (\ref{3.13}) subject to boundary conditions (\ref{3.15}) (for $k=0$)  and the first of the conditions (\ref{3.16}) represent
the inviscid eigenvalue problem that arises in linear stability analysis of the inviscid flow
\begin{equation}
u=\frac{1}{r}, \quad V=\frac{\gamma_1}{r}. \label{3.18}
\end{equation}
This inviscid problem had been studied in detail in \cite{IM2013}.
Note that at this stage we cannot satisfy the second of the conditions (\ref{3.16}), and that is why a boundary layer at $r=a$ is needed.

To derive equations for the boundary layer part of the expansion, we substitute (\ref{3.8}) and (\ref{3.9})
into Eq. (\ref{3.11}) and take into account that $\hat{\psi}_{k}$ ($k=0,1,\dots$)
satisfy (\ref{3.13}), (\ref{3.14}). Then we make the change of variable
$r=a(1-R^{-1} \, \eta)$, expand every function of $a(1-R^{-1} \, \eta)$ in Taylor's series at $R^{-1}=0$
and, finally, collect terms of the equal powers of $R^{-1}$. At leading order, we obtain
\begin{equation}
- \pr_{\eta}^3 {\phi}_{0}=  \pr_{\eta}^4 {\phi}_{0}.  \label{3.19}
\end{equation}
Evidently, neither $\sigma_0$ nor boundary layer parts of the basic velocity or vorticity enter the leading-order equation.
Equation (\ref{3.19}) should be solved subject to the condition of decay at infinity (in variable $\eta$) and the second condition
(\ref{3.16}) that can be written as
\begin{equation}
\phi'_{0}(0)=\hat{\psi}'_{0}(a). \label{3.20}
\end{equation}
The solution of
(\ref{3.19}), (\ref{3.20}) is
\begin{equation}
\phi_{0}(\eta)=- \hat{\psi}'_{0}(a)e^{-\eta}. \label{3.21}
\end{equation}
Note that the boundary layer does not affect the leading order eigenvalue $\sigma_{0}$ which is completely determined by
the inviscid problem. It can be shown that the boundary layer will
affect the first order correction $\sigma_{1}$, but we will not compute it here.

Thus, we have shown that in the limit $R\to\infty$, the linearised stability problem for the viscous flow
(\ref{5}) with $\beta=1$ reduces to the inviscid stability problem for the irrotational flow (\ref{3.18}).

\subsubsection{Converging flow ($\beta=-1$)}

For the converging flow ($\beta=-1$), the situation is similar except that now the boundary layer is near the inner cylinder. The asymptotic
expansion has the form
\begin{eqnarray}
&& \sigma=\sigma_{0}+R^{-1}\sigma_{1}+R^{-2}\sigma_{1}+\dots ,  \label{3.22} \\
&& \hat{\psi}=\hat{\psi}_{0}(r)+R^{-1} [\hat{\psi}_{1}(r)+\phi_{0}(\xi)]+R^{-2} [\hat{\psi}_{2}(r)+\phi_{1}(\xi)]+\dots,  \label{3.23}
\end{eqnarray}
where $\xi=R(r-1)$ is the boundary layer variable (the same as that introduced in section 2). Again, the boundary layer part is assumed to
rapidly decay outside thin boundary layer near $r=1$:
\begin{equation}
\phi_{k}(\xi) =o(\xi^{-s}) \quad {\rm as} \quad \xi\to\infty ,    \label{3.24}
\end{equation}
for every $s>0$ and for each $k=0,1,\dots$

Substituting asymptotic formulae (\ref{12}) and (\ref{13}) into Eq. (\ref{3.6}), we get
\begin{equation}
\left(\sigma +  \frac{in \gamma_2}{r^2} - \frac{in}{r} \, g(\xi)
+\frac{1}{r} \, D \right) L \hat{\psi}
-  R^2 \, \frac{in}{r} \, g(\xi) \, \hat{\psi} = R^{-1}L^2 \hat{\psi}    \label{3.25}
\end{equation}
where
\[
g(\xi)=\left(\gamma_1 - \gamma_2\right) \, e^{-\xi} \left[1+O\left(R^{-1}\right)\right].
\]
A repetition of the analysis of the preceding subsection leads to the conclusion that the leading-order eigenvalue
is a solution of the eigenvalue problem
\begin{eqnarray}
&&\left(\sigma_{0} +  \frac{in\gamma_2}{r^2} - \frac{1}{r} \, \pr_{r} \right)L \hat{\psi}_{0}= 0, \label{3.26} \\
&&\hat{\psi}_{0}(a) = 0, \quad \hat{\psi}'_{0}(a)=0 , \quad \hat{\psi}_{0}(1)=0, \label{3.27}
\end{eqnarray}
which arises in the linear stability analysis of the inviscid flow
\begin{equation}
u=-\frac{1}{r}, \quad V=\frac{\gamma_2}{r} \label{3.28}
\end{equation}
and which had been studied by \cite{IM2013}. As in the preceding subsection, there is a boundary layer which, for the converging flow, is
near the inner cylinder, but it does not affect the leading eigenvalue and will not be discussed here.

Thus, we have shown that in the limit $R\to\infty$, the linearised stability problem for the viscous flow
(\ref{5}) with $\beta=-1$ reduces to the inviscid stability problem for the irrotational flow (\ref{3.28}).

The above analysis and the results of \cite{IM2013} imply that for sufficiently high Reynolds numbers $R$, the steady flow (\ref{5})
is unstable to small two-dimensional perturbations in a wide range of the parameters of the problem (see \cite{IM2013}).


\subsection{Numerical results}

Numerical calculations of the eigenvalue problem (\ref{3.6}), (\ref{3.7}) show the following:
(i) the purely radial flow ($\gamma_1=\gamma_2=0$) is stable (i.e. $\Real(\sigma)< 0$) for both $\beta=1$ and $\beta=-1$, for any $a>1$ and $R>0$ and
for all azimuthal modes;
(ii) the diverging flow ($\beta=1$) with $\gamma_1=0$, i.e. with zero azimuthal velocity at the inlet, is stable for all
$\gamma_2$, $a$ and $R$; and
(iii) the converging flow ($\beta=-1$) with $\gamma_2=0$, i.e. again with zero azimuthal velocity at the inlet, is stable for all
$\gamma_1$, $a$ and $R$.

Below we describe the results for the diverging and converging flows separately.

\subsubsection{Diverging flow ($\beta=1$)}

\begin{figure}
\begin{center}
\includegraphics*[height=8cm]{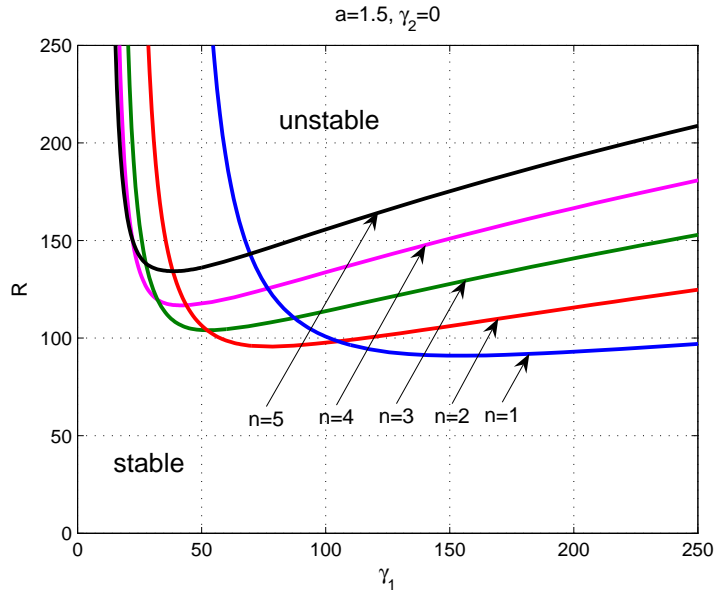}
\end{center}
\caption{Diverging flow: neutral curves for $a=1.5$, $\gamma_2=0$ and $n=1, \dots, 5$. The region above each curve is
where the corresponding mode is unstable.}
\label{fig2}
\end{figure}

\begin{figure}
\begin{center}
\includegraphics*[height=8cm]{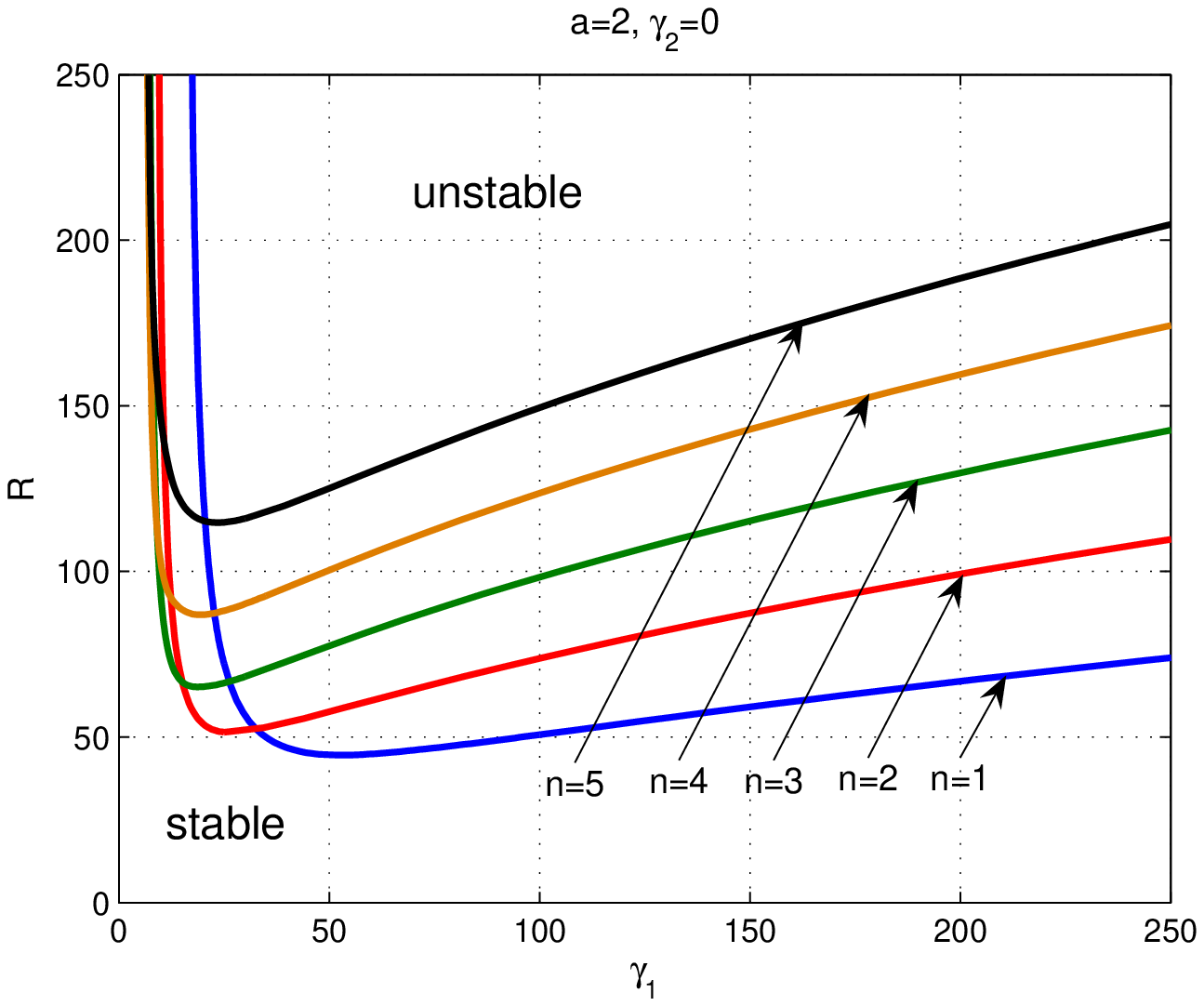}
\end{center}
\caption{Diverging flow: neutral curves for $a=2$, $\gamma_2=0$ and $n=1, \dots, 5$. The region above each curve is
where the corresponding mode is unstable.}
\label{fig3}
\end{figure}

\begin{figure}
\begin{center}
\includegraphics*[height=8cm]{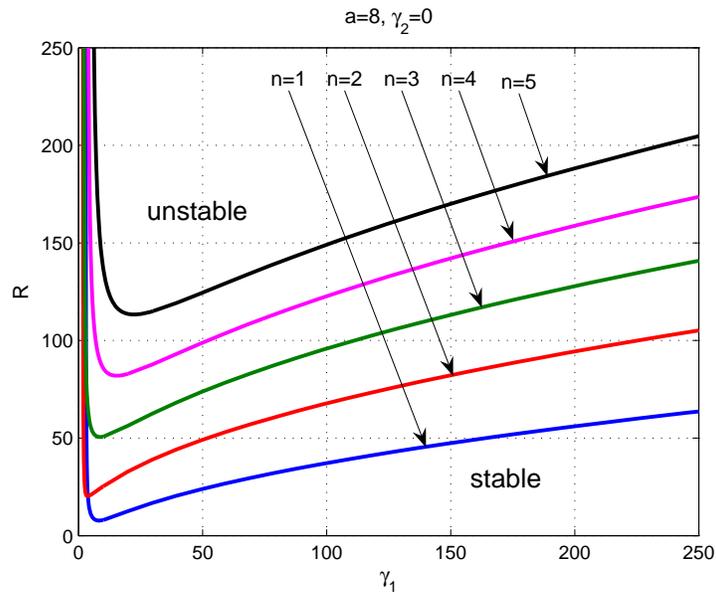}
\end{center}
\caption{Diverging flow: neutral curves for $a=8$, $\gamma_2=0$ and $n=1, \dots, 5$. The region above each curve is
where the corresponding mode is unstable.}
\label{fig4}
\end{figure}

We have computed neutral curves on the $(\gamma_1,R)$ plane for first 5 azimuthal modes ($n=1, \dots,5$) and
for various values of $a$ and $\gamma_2$. Neutral curves for $\gamma_2=0$ and $a=1.5$, $a=2$ and $a=8$ are shown in Figures \ref{fig2},
\ref{fig3} and \ref{fig4} respectively.
Figures \ref{fig2}--\ref{fig4} show that, for all values of $a$, the neutral curves look qualitatively similar.
The curves shift down when the radius of the outer cylinder, $a$, is increased. This means that the critical Reynolds number, $R_c$,
defined as a value of $R$ which separates the region of stability and instability, decreases with $a$.
Each neutral curve approaches a vertical asymptote, $\gamma_1=\gamma_*(a,n)$, where $\gamma_*(a,n)$ is a critical value of
$\gamma_1$ separating stable and unstable inviscid flows (\ref{3.18}).

\begin{figure}
\begin{center}
\includegraphics*[height=8cm]{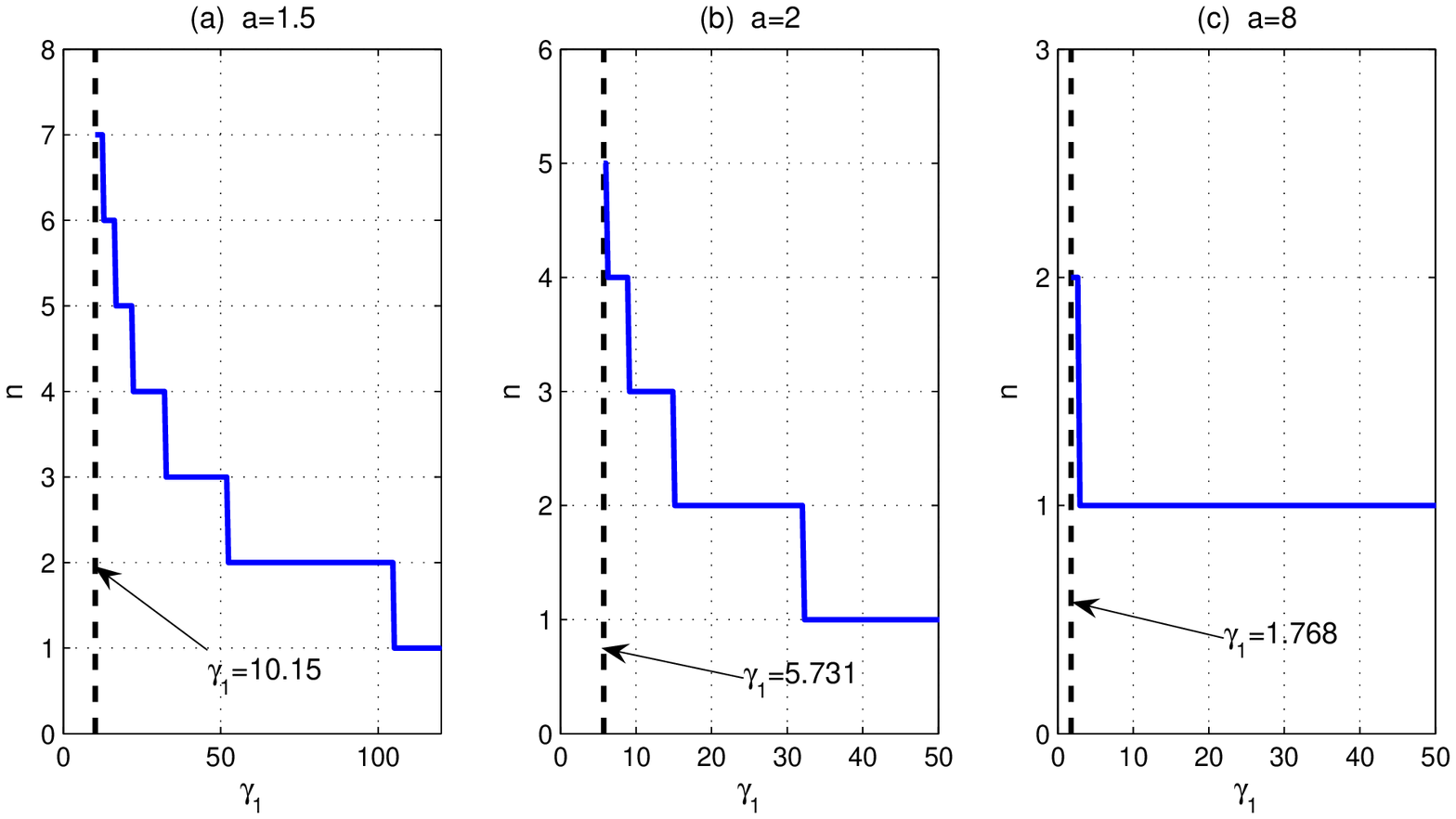}
\end{center}
\caption{Diverging flow: azimuthal wavenumber, $n$, of the most unstable mode as a
function of $\gamma_1$ for $\gamma_2=0$. (a) $a=1.5$; (b) $a=2$; (a) $a=8$. }
\label{fig5}
\end{figure}

The azimuthal wave number $n$ of the most unstable mode (i.e. the mode that becomes unstable first when $R$ is increased from zero)
depends on $\gamma_1$. The wave numbers $n$ of the most unstable modes versus $\gamma_1$ for $a=1.5$, $a=2$ and $a=8$ are shown
Fig. \ref{fig5}.
The jumps in $n$ correspond to the intersection points of the neutral curves in Figures \ref{fig2}--\ref{fig4}.
Dashed vertical lines represent $\gamma_1=\gamma_*(a,n)$. Clearly, the number of different azimuthal modes which can be
the most unstable mode depend on $a$. For $a=1.5$, each mode with $n$ ranging from $1$ to $7$ can be the most unstable one
(depending on $\gamma_1$). For $a=2$, each mode with $n=1, \dots, 5$ can be the most unstable one. For $a=8$ (and for larger
values of $a$), only one of two modes with $n=1$ and $n=2$ can be the most unstable one.

\begin{figure}
\begin{center}
\includegraphics*[height=8cm]{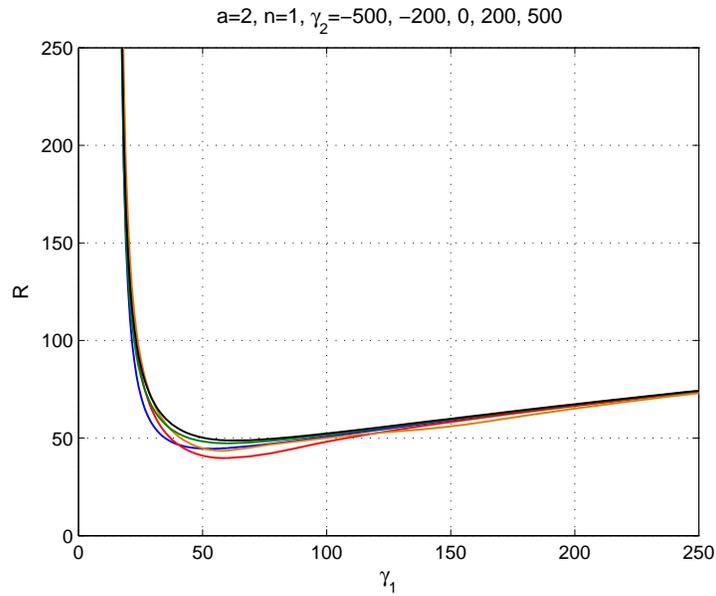}
\end{center}
\caption{Diverging flow: neutral curves for $n=1$, $a=2$ and $\gamma_2=-500,-200,0,200,500$.}
\label{fig6}
\end{figure}

The neutral curves also depend on parameter $\gamma_2$. However, this dependence is very weak, and neutral curves for
finite nonzero values of $\gamma_2$ remain qualitatively the same as those shown in Figs. \ref{fig2}--\ref{fig4}. This fact
is illustrated in Fig. \ref{fig6} where neural curves
for $n=1$, $a=2$ and several values of $\gamma_2$ are presented. Figure \ref{fig6} shows that for quite a wide range of $\gamma_2$ ($\gamma_2\in[-500,500]$) the neutral curves are very close to the one shown in Fig. \ref{fig2}.

\subsubsection{Converging flow ($\beta=-1$)}

\begin{figure}
\begin{center}
\includegraphics*[height=8cm]{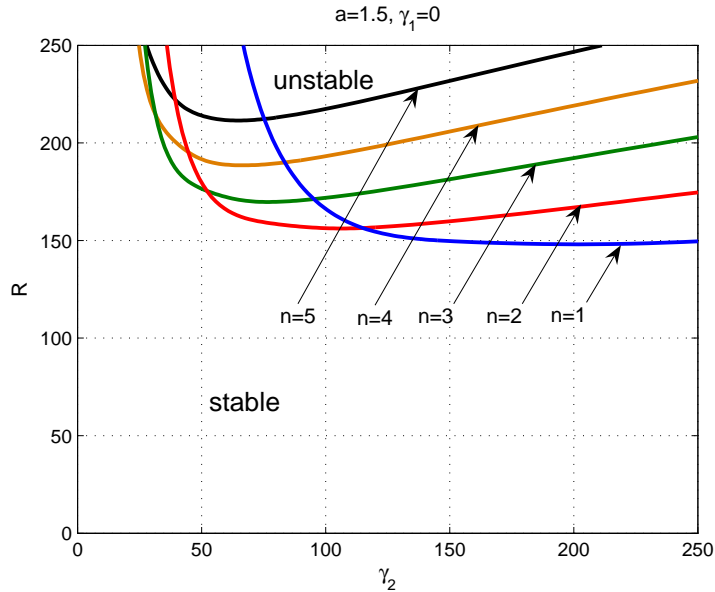}
\end{center}
\caption{Converging flow: neutral curves for $a=1.5$, $\gamma_1=0$ and $n=1, \dots, 5$. The region above each curve is
where the corresponding mode is unstable.}
\label{fig7}
\end{figure}

\begin{figure}
\begin{center}
\includegraphics*[height=8cm]{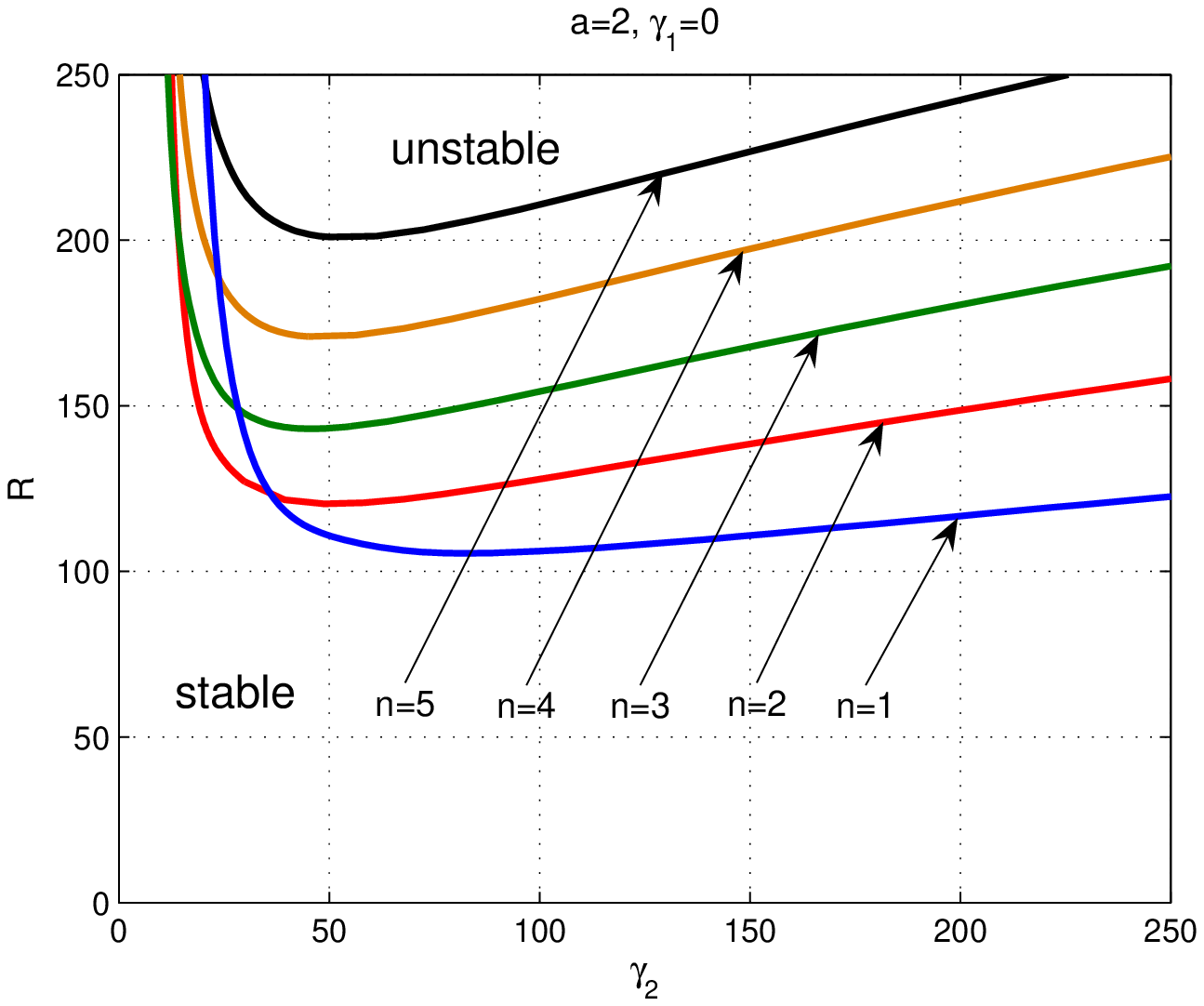}
\end{center}
\caption{Converging flow: neutral curves for $a=2$, $\gamma_1=0$ and $n=1, \dots, 5$. The region above each curve is
where the corresponding mode is unstable.}
\label{fig8}
\end{figure}

\begin{figure}
\begin{center}
\includegraphics*[height=8cm]{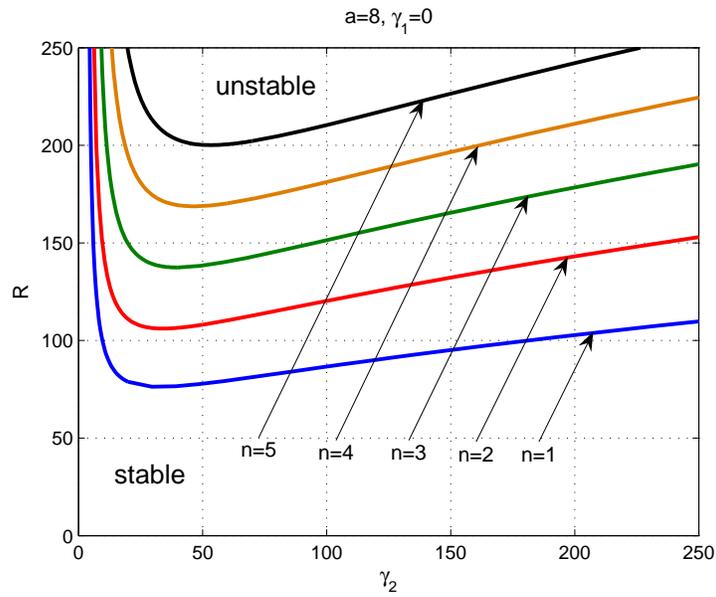}
\end{center}
\caption{Converging flow: neutral curves for $a=8$, $\gamma_1=0$ and $n=1, \dots, 5$. The region above each curve is
where the corresponding mode is unstable.}
\label{fig9}
\end{figure}

\begin{figure}
\begin{center}
\includegraphics*[height=8cm]{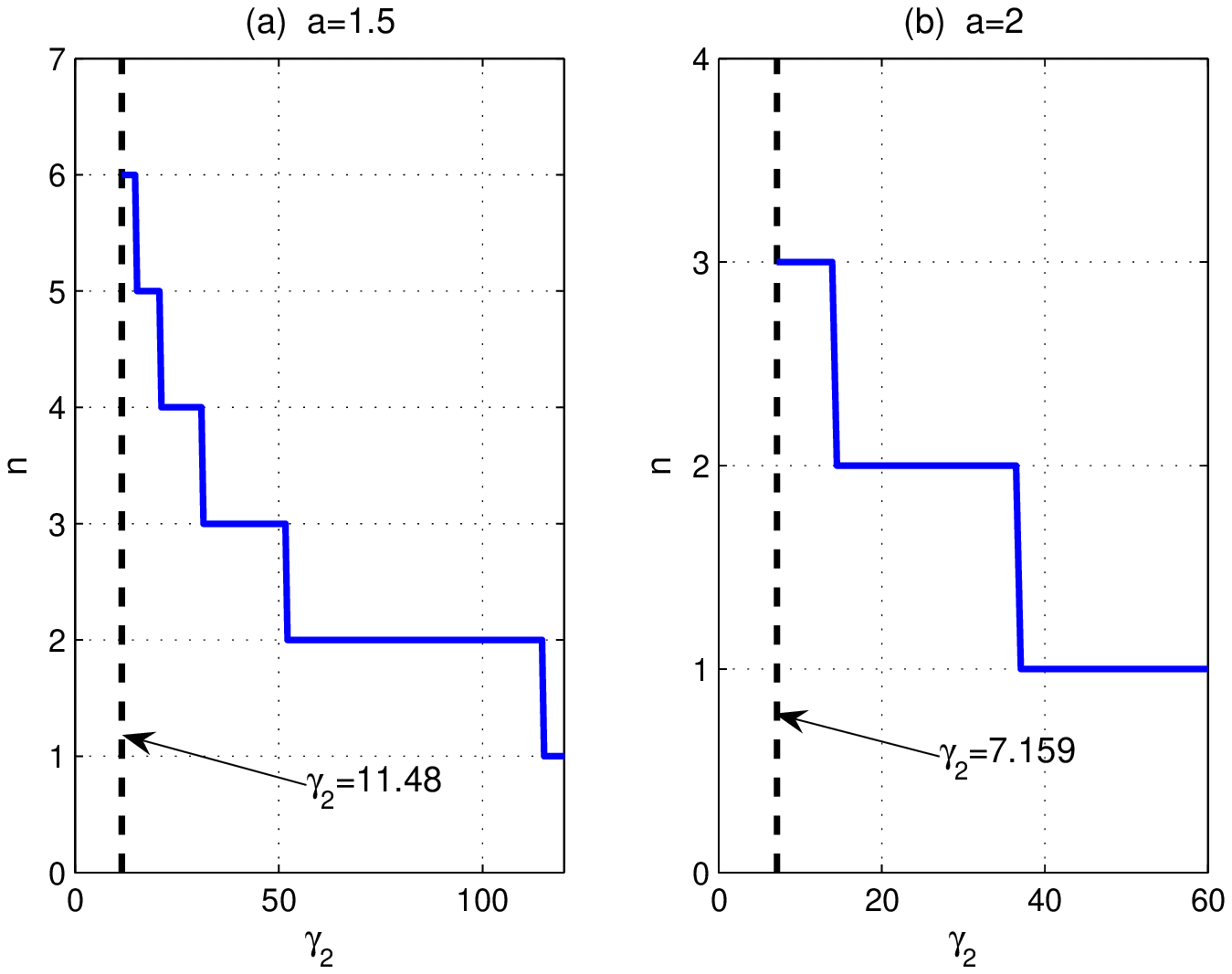}
\end{center}
\caption{Converging flow: azimuthal wavenumber $n$ of the most unstable mode as a
function of $\gamma_2$ for $\gamma_1=0$. (a) $a=1.5$; (b) $a=2$. }
\label{fig10}
\end{figure}

The most apparent difference between the diverging and converging flows is that the stability properties
of the latter are determined by parameter $\gamma_2$ rather than $\gamma_1$. This is a manifestation of the general
property of flows with non-zero normal velocity at the boundary: the inflow part of the boundary
plays much more important role than the outflow part. In other words,
it matters what azimuthal velocity fluid particles have when they enter the flow domain,
but their velocity at the outlet is far less important.
Neutral curves on the $(\gamma_2,R)$ plane for first 5 azimuthal modes ($n=1, \dots,5$) and for $\gamma_1=0$
are shown in Figures \ref{fig7}--\ref{fig9}.
These curves are qualitatively similar to what we had for the diverging flow (apart from the fact
that here we use the $(\gamma_2,R)$ plane rather than the $(\gamma_1,R)$ plane).
As before, each neutral curve approaches a vertical asymptote, $\gamma_1=\gamma_*(a,n)$, where $\gamma_*(a,n)$ is a critical value of
$\gamma_1$ separating stable and unstable inviscid flows (\ref{3.28}).
The azimuthal wave number $n$ of the most unstable mode
depends on $\gamma_2$. The wave numbers $n$ of the most unstable modes versus $\gamma_1$ for $a=1.5$ and $a=2$ are shown
Fig. \ref{fig10}.
The jumps in $n$ correspond to the intersection points of the neutral curves in Figures \ref{fig7} and \ref{fig8}.
Dashed vertical lines represent $\gamma_1=\gamma_*(a,n)$. For $a=1.5$, each mode with $n=1,\dots,6$ can be the most unstable one
(depending on $\gamma_2$). For $a=2$, only modes with $n=1,2,3$ can be the most unstable one. For $a=8$ (and larger values of $a$),
the most unstable mode for all $\gamma_2$ is the mode with $n=1$.

\begin{figure}
\begin{center}
\includegraphics*[height=8cm]{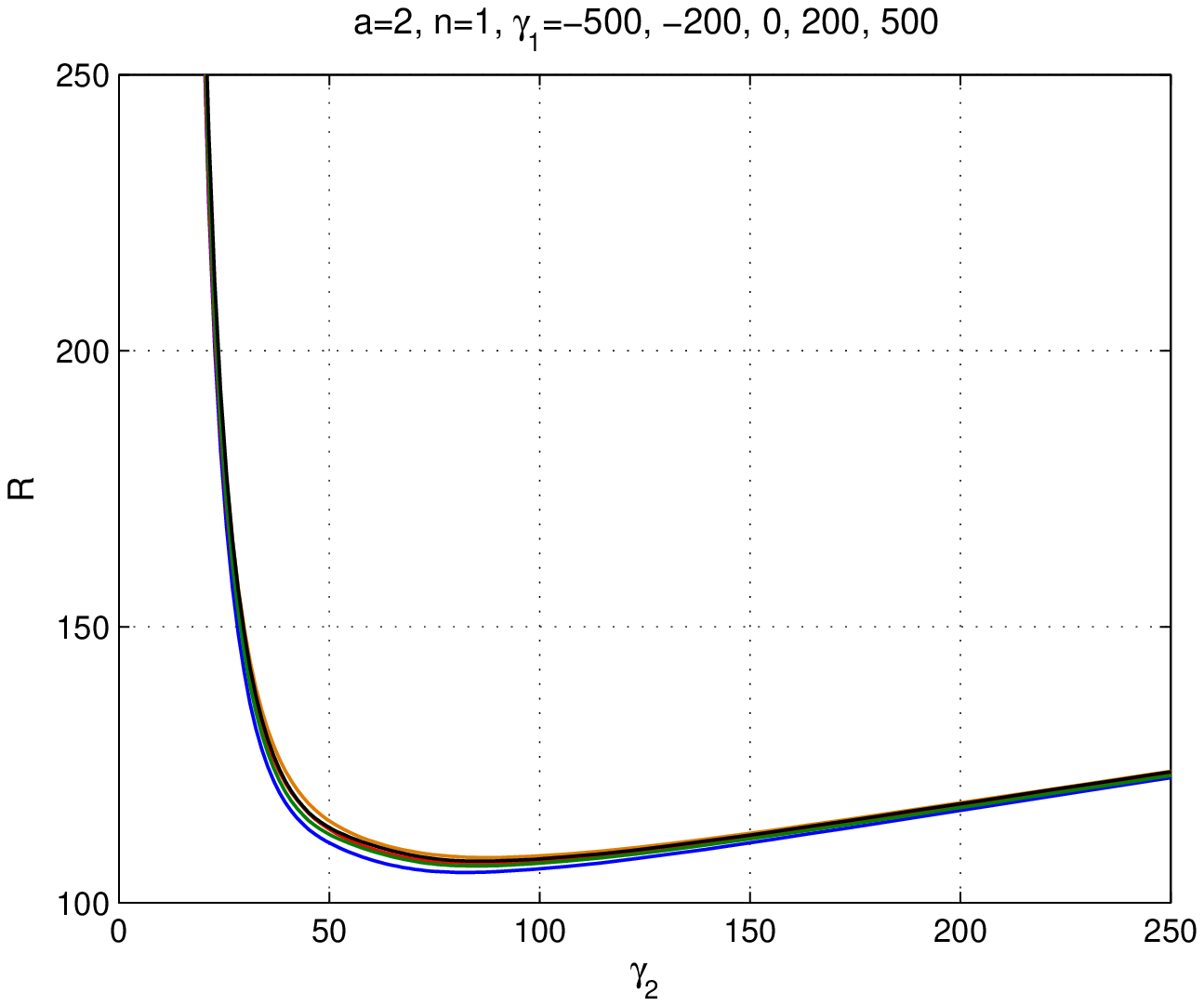}
\end{center}
\caption{Converging flow: Neutral curves for $n=1$, $a=2$ and $\gamma_1=-500,-200,0,200,500$.}
\label{fig11}
\end{figure}

The neutral curves also depend on parameter $\gamma_1$, but again this dependence is weak, as illustrated in Fig. \ref{fig11} where neural curves
for $n=1$, $a=2$ and several values of $\gamma_1$ are shown.

\section{Conclusions}

We have shown that the instability of a simple steady inviscid flow found in \cite{IM2013} also occurs for a wide class of viscous flows
between rotating porous cylinders provided that the radial Reynolds number is sufficiently high.

Our calculations show that both purely radial and purely azimuthal flows are always stable to two-dimensional perturbations.
If we add a radial component to a sufficiently strong purely azimuthal basic flow and gradually increase its magnitude, the flow becomes unstable when
the Reynolds number (based on the radial velocity) passes through a certain critical value which depends on the ratio of the radii of the cylinders and on their angular velocities. In the case of the diverging flow, this critical Reynolds number strongly depends
on the angular velocity of the inner cylinder and is almost independent of the angular velocity of the outer cylinder.
For the converging flow, the situation is opposite: the critical Reynolds number is strongly affected by the angular velocity
of the outer cylinder and is almost independent of the angular velocity of the outer cylinder.
These facts indicate that properties of viscous flows through a given domain are almost unaffected by the flow outlet, but the conditions
at the inflow part of the boundary can affect the flow in the whole domain. What is interesting is that for viscous flows through finite channels or pipes, the situation appears to be different: the effect of the inlet on the flow downstream weakens with the distance
due to viscosity, so that the details of the flow near the inlet are not important for the flow far from it. The results presented here show that
there are viscous flows for which the effect of the flow near inlet remains strong in the whole flow domain.

In \cite{IM2013}, it was noted that
the most unexpected result of that study was that not only the diverging flow but
also the converging flow are unstable and that
this  fact differs from known results on the viscous stability of the Jeffery-Hamel flow and, especially, the vortex source/sink flow which
appear to suggest that converging flows are always stable while diverging flows may be unstable
(\cite{Goldshtik1991,Shtern,Drazin1998}). The present study confirms that both diverging and converging viscous flows can be unstable,
provided that the azimuthal velocity at the inlet is sufficiently large. However, the neutral curves in Figures
\ref{fig2}--\ref{fig4} and \ref{fig7}--\ref{fig9} show that, in all cases, the critical Reynolds numbers for converging flows
are considerably higher than those for diverging flows. In this sense, diverging flows are `more unstable' than converging flows.

One of the most interesting open question is the stability of the steady flow (\ref{5}) to three-dimensional perturbation, and this is a topic of a continuing investigation.

The results presented here are mainly of theoretical interest. However
a further development of the theory may lead to results applicable to
the process of dynamic filtration using rotating filters (e.g. \cite{Wron'ski}).

We are grateful to Professor V. A. Vladimirov for helpful discussions.

\section{Appendix A}

Here we show that if $n=0$, then $\Real(\sigma) < 0$, i.e. the mode with $n=0$ cannot be unstable.
Indeed,
Eq. (\ref{3.4}) for $n=0$ and the boundary conditions for $\hat{u}$ imply that $\hat{u}=0$.
Equation (\ref{3.3}) simplifies to
\[
\left(\sigma + \frac{1}{r} \, \pr_{r} \right) \hat{v}
+\frac{1}{r^2} \, \hat{v}  =
\frac{1}{R} \left(L_0 \hat{v} -\frac{\hat{v}}{r^2}\right), \quad L_0 \equiv \frac{d^2}{dr^2} + \frac{1}{r} \, \frac{d}{dr}.
\]
We multiply this equation by $r \hat{v}^{*}$ where $\hat{v}^{*}$ is the complex conjugate of $\hat{v}$ and integrate
in $r$ from $1$ ro $a$. This yields
\[
\int\limits_{0}^{a}\left(\sigma r \vert\hat{v}\vert^2 + \hat{v}^{*}\hat{v}_{r}+\frac{1}{r}\vert\hat{v}\vert^2\right) \, dr =
\frac{1}{R} \int\limits_{0}^{a}\left(\hat{v}^{*}\pr_{r}(r\hat{v}_r)- \frac{1}{r}\vert\hat{v}\vert^2 \right) \, dr
\]
Integration by parts and some manipulations reduce this to
\[
\sigma \int\limits_{0}^{a}\vert\hat{v}\vert^2 \, r \, dr +
\int\limits_{0}^{a} \hat{v}^{*}\hat{v}_{r} \, r \, dr
= -
\left(1+\frac{1}{R}\right) \int\limits_{0}^{a}\frac{1}{r^2}\vert\hat{v}\vert^2 \, r \, dr
-\int\limits_{0}^{a}\vert\hat{v}_r\vert^2\, r \, dr.
\]
Taking the real part of this equation, we obtain
\[
\Real (\sigma) \int\limits_{0}^{a}\vert\hat{v}\vert^2 \, r \, dr = -
\left(1+\frac{1}{R}\right) \int\limits_{0}^{a}\frac{1}{r^2}\vert\hat{v}\vert^2 \, r \, dr
-\int\limits_{0}^{a}\vert\hat{v}_r\vert^2\, r \, dr.
\]
It follows that $\Real (\sigma)<0$.

\bibliographystyle{jfm}


\begin{thebibliography}{100}


\bibitem[Bahl (1970)]{Bahl}
{Bahl, S. K. 1970 Stability of viscous flow between two concentric rotating
porous cylinders.
{\em Def. Sci. J.}, {\bf 20}(3), 89--96.}

\bibitem[Beavers \& Joseph (1967)]{Joseph}
{Beavers, G. S. \& Joseph, D. D. 1967 Boundary conditions at a naturally permeable wall.
{\em J. Fluid Mech.}, {\bf 30}(1), 197-207.}

\bibitem[Chang \& Sartory (1967)]{Chang}
{Chang, S. \& Sartory, W. K. 1967
Hydromagnetic stability of dissipative flow between rotating
permeable cylinders.
{\em J. Fluid Mech.}, {\bf 27}, 65--79.}






\bibitem[Goldshtik et al (1991)]{Goldshtik1991}
{Goldshtik, M.,  Hussain, F.  \& Shtern, V. 1991 Symmetry breaking in vortex-source and Jeffery—Hamel flows.
{\em J. Fluid Mech.}, {\bf 232}, 521--566.}


\bibitem[Ilin (2008)]{Ilin2008}
{Ilin, K. 2008 Viscous boundary layers in flows through a domain
with permeable boundary. {\em Eur. J. Mech. B/Fluids}, {\bf 27}, 514–-538.}

\bibitem[Ilin \& Morgulis (2013)]{IM2013}
{Ilin, K. \& Morgulis, A. 2013 Instability of an inviscid flow between porous cylinders with radial flow. {\em J. Fluid Mech.}
{\bf 730}, 364--378.}


\bibitem[Kolesov \& Shapakidze (1999)]{Kolesov}
{Kolesov, V. \& Shapakidze, L. 1999 On oscillatory modes in viscous incompressible liquid flows between two counter-rotating permeable cylinders.
In: {\em Trends in Applications of Mathematics to Mechanics} (ed. G. Iooss, O. Gues \& A Nouri), pp. 221--227. Chapman and Hall/CRC.}

\bibitem[Kolyshkin \& Vaillancourt (1997)]{Kolyshkin}
{Kolyshkin, A. A. \& Vaillancourt, R. 1997
Convective instability boundary of Couette flow between rotating porous cylinders
with axial and radial flows.
{\em Phys. Fluids}, {\bf 9}, 9, 910--918.}

\bibitem[McAlpine \& Drazin (1998)]{Drazin1998}
{McAlpine, A. \& Drazin, P. G. 1998 On the spatio-temporal development of small perturbations of Jeffery-Hamel flows.
{\em Fluid Dynamics Research}, {\bf 22}(3), 123–-138.}

\bibitem[Min \& Lueptow (1994)]{Min}
{Min, K. \& Lueptow, R. M. 1994
Hydrodynamic stability of viscous flow between rotating porous cylinders
with radial flow. {\em Phys. Fluids}, {\bf 6}, 144--151.}



\bibitem[Shtern \& Hussain (1993)]{Shtern}
{Shtern, V. \& Hussain, F. 1993 Azimuthal instability of divergent flows.
{\em J. Fluid Mech.}, {\bf 256}, 535--560.}

\bibitem[Trefethen (2000)]{Trefethen}	
Trefethen, L. N. 2000 \textit{Spectral methods in MATLAB.} Vol. 10. Siam.

\bibitem[Temam \& Wang (2000)]{Temam}
 {Temam, R. \& Wang, X.} 2000 Remarks on the Prandtl equation for a permeable wall.
 {\em Z. Angew. Math. Mech.\/} {\bf 80}, 835--843.

\bibitem[Wron'ski et al (1989)]{Wron'ski}
 {Wron'ski, S., Molga, E. \& Rudniak, L.} 1989 Dynamic filtration in biotechnology.
 {\em Bioprocess Engineering\/} {\bf 4}(3), 99--104.

\bibitem[Yudovich (2001)]{Yudovich2001}
{Yudovich, V.~I.} 2001 Rotationally symmetric flows of incompressible
fluid through an annulus. Parts I and II. {\em Preprints VINITI\/}
no.~1862-B01 and no.~1843-B01 [in Russian].











\end{thebibliography}

\end{document}